\documentclass{article}
\usepackage{spconf,amsmath,graphicx,verbatim}
\usepackage{subcaption, comment, multirow, multicol, makecell, amssymb, bm, textcomp, kotex, arydshln}
\usepackage{adjustbox}
\usepackage{mathtools}
\usepackage{cite, url}
\usepackage{tabularx, makecell}
\usepackage{hyperref}
\usepackage{xcolor}
\hypersetup{
    colorlinks=true,
    linkcolor=purple,
    filecolor=magenta,      
    urlcolor=cyan,
}
\newcolumntype{Y}{>{\centering\arraybackslash}X}
\usepackage[subtle]{savetrees}

\title{Integrated Parameter-Efficient Tuning for General-Purpose Audio Models}
\name{Ju-ho Kim$^*$\thanks{$^*$Equal contribution}, Jungwoo Heo$^*$, Hyun-seo Shin, Chan-yeong Lim and Ha-Jin Yu\sthanks{$^\dag$Corresponding author}\thanks{This research was supported by Basic Science Research Program through the National Research Foundation of Korea (NRF) funded by the Ministry of Science, ICT \& Future Planning (2020R1A2C1007081)}}
\address{School of Computer Science, University of Seoul}

\begin{document}

\ninept
\maketitle
\begin{abstract}
The advent of hyper-scale and general-purpose pre-trained models is shifting the paradigm of building task-specific models for target tasks. 
In the field of audio research, task-agnostic pre-trained models with high transferability and adaptability have achieved state-of-the-art performances through fine-tuning for downstream tasks. 
Nevertheless, re-training all the parameters of these massive models entails an enormous amount of time and cost, along with a huge carbon footprint. 
To overcome these limitations, the present study explores and applies efficient transfer learning methods in the audio domain. 
We also propose an integrated parameter-efficient tuning (IPET) framework by aggregating the embedding prompt (a prompt-based learning approach), and the adapter (an effective transfer learning method). 
We demonstrate the efficacy of the proposed framework using two backbone pre-trained audio models with different characteristics: the audio spectrogram transformer and wav2vec 2.0. 
The proposed IPET framework exhibits remarkable performance compared to fine-tuning method with fewer trainable parameters in four downstream tasks: sound event classification, music genre classification, keyword spotting, and speaker verification. 
Furthermore, the authors identify and analyze the shortcomings of the IPET framework, providing lessons and research directions for parameter efficient tuning in the audio domain. 
\end{abstract}
\begin{keywords}
general-purpose audio model, audio spectrogram transformer, wav2vec2.0, transfer learning, prompt-based learning
\end{keywords}

\section{Introduction}
\label{sec:intro}
From natural language processing (NLP) to computer vision and audio signal processing, task-agnostic pre-trained models (PM) are attracting researchers' attention due to their versatile potential for use in a variety of downstream tasks \cite{radford2019language, DBLP:conf/iclr/DosovitskiyB0WZ21, gong2022ssast}. 
An architectural trend of PMs is to significantly increase the magnitude and complexity based on transformers \cite{vaswani2017attention} to solve challenging and complicated tasks \cite{radford2019language, DBLP:conf/naacl/DevlinCLT19}. 
For various downstream tasks, the adaptation of hyper-scale PM surpasses the conventional task-specific model approaches, but entails high memory and costs for fine-tuning. 
In addition, it is not certain that updating the entire parameters is the optimal scheme to leverage the PM's knowledge \cite{DBLP:journals/corr/abs-2103-10385}.

In NLP, recent research has shown that general-purpose language PM could be suited to downstream tasks by utilizing prompts without fine-tuning \cite{schick-schutze-2021-exploiting}. 
Prompting is a method of reconstructing data to control the output of PM \cite{DBLP:journals/corr/abs-2107-13586}. 
As an example, for the sentiment classification task of a restaurant review, there is an input of \textit{``Best Korean BBQ of my life!"}. 
The language PM cannot deduce the sentiment label for this sentence alone without recognizing the task. 
Here, a prompt, \textit{``The restaurant was} [mask]\textit{."}, is appended to convey information about the target task to the PM. 
For the prompted sentence, the PM can directly derive the label (e.g., \textit{``great"} or \textit{``awesome"}) to fill the mask by understanding the downstream task without re-training. 
As such, the prompt technique can successfully adapt PM to various sub-tasks, but manually navigating to an appropriate prompt is expensive and difficult to apply to image or audio domain data \cite{DBLP:journals/corr/abs-2103-10385}. 
To overcome these drawbacks, several researchers have devised schemes to explore learnable prompts in continuous space beyond discrete space. 
These approaches, called prompt-based learning, which add trainable parameters to the data level \cite{zhou2022conditional, bahng2022exploring} or embedding level \cite{DBLP:journals/corr/abs-2103-10385, DBLP:journals/corr/abs-2203-12119}, have proven more effective than conventional discrete prompts and can be applied to other domain data. 

Meanwhile, computationally efficient transfer learning is also being investigated to better utilize PM. 
Linear probing is a straightforward transfer learning method that trains only the last linear layer specialized for sub-tasks, while keeping the PM's weights fixed. 
It is simple and intuitive but sub-optimal, and tends to miss opportunities to learn task-specific features \cite{he2022masked}. 
The adapter \cite{rebuffi2017learning} proposed to overcome this shortcoming is a representative parameter-efficient transfer learning strategy. 
This method adds a few linear layers with activation function to each transformer block of the PMs. 
In this modified PM, only the adapter's parameters are trained for the target task, while the weight of the original PM is frozen. 
The adapter approaches have shown enhanced results by changing the model's behavior to parameter-efficiently fit sub-tasks \cite{houlsby2019parameter}.

Prompt-based learning and adapter techniques have different mechanisms and principles for using PMs, but similar goals (i.e., to adjust the model's results for downstream tasks without modifying any of the original parameters). 
Leveraged on this objective, each paradigm has been successfully introduced into the audio domain, yielding promising results for transfer learning scenarios \cite{chang2022exploration, thomas2022efficient}. 
However, these parameter-efficient tuning approaches have been individually studied, despite each method being related. 
We propose an integrated parameter-efficient tuning (IPET) framework for the audio domain by fusing effective and efficient prompt-based learning and adapter strategies. 
The IPET framework can be directly applied to general-purpose audio PMs in an end-to-end manner with memory and cost efficiency for various audio downstream tasks.

As the backbone PM of the proposed framework, we employed the audio spectrogram transformer (AST) \cite{gong2021ast} trained in a supervised learning manner with the general data of the audio domain, and wav2vec2.0 (W2V2) \cite{baevski2020wav2vec} trained in a self-supervised learning scheme to extract general features of the speech data. 
The prompt-based learning is applied to the input (Mel-spectrogram or raw waveform) or embedding space to shift the model's data distribution, and the adapter is introduced to the block of the backbone model to alter the model's behavior. 
The proposed IPET framework combines both techniques simultaneously to coordinate the PM for downstream tasks. 
Experimental results exhibit that the IPET framework achieved superior performance using a limited number of learnable parameters under specific conditions compared to the independent tuning approaches. 
Simultaneously, we also identified and analyzed the limitations of the proposed framework. 
This paper discusses the lessons and future works for parameter-efficient tuning in the audio domain based on observations of experiment results. 

\section{Related work}
\label{sec:rw}
Recently, several studies have been exploring how to efficiently exploit PM in audio domain. 
Audio-language PMs have performed zero-shot and few-shot evaluations for audio sub-tasks by applying fixed natural language prompts to text input \cite{DBLP:journals/corr/abs-2206-04769, gao2022wavprompt}. 
Chang \textit{et al.} \cite{chang2022exploration} solved speech processing tasks by introducing the learnable embedding prompts to the language model. 
While these studies are relevant to us in terms of the adoption of prompts for audio downstream tasks, the novelty of our study lies in exploring prompt-based learning directly at the input and embedding levels of audio models, excluding additional language models. 

On the other hand, in the field of parameter-efficient transfer learning, Gerczuk \textit{et al.} \cite{gerczuk2021emonet} performed domain adaptation in the emotion recognition task using the CNN-based adapter module. 
In addition, Thomas \textit{et al.} \cite{thomas2022efficient} transferred the self-supervised PM to the speech recognition model using the adapter. 
This literature further explores a framework for effectively adapting general-purpose audio PMs to various audio down-stream tasks by combining prompt-based learning and parameter-efficient transfer learning schemes. 

\begin{figure}[th!]
\begin{center}
    \centering
    \includegraphics[width=\linewidth]{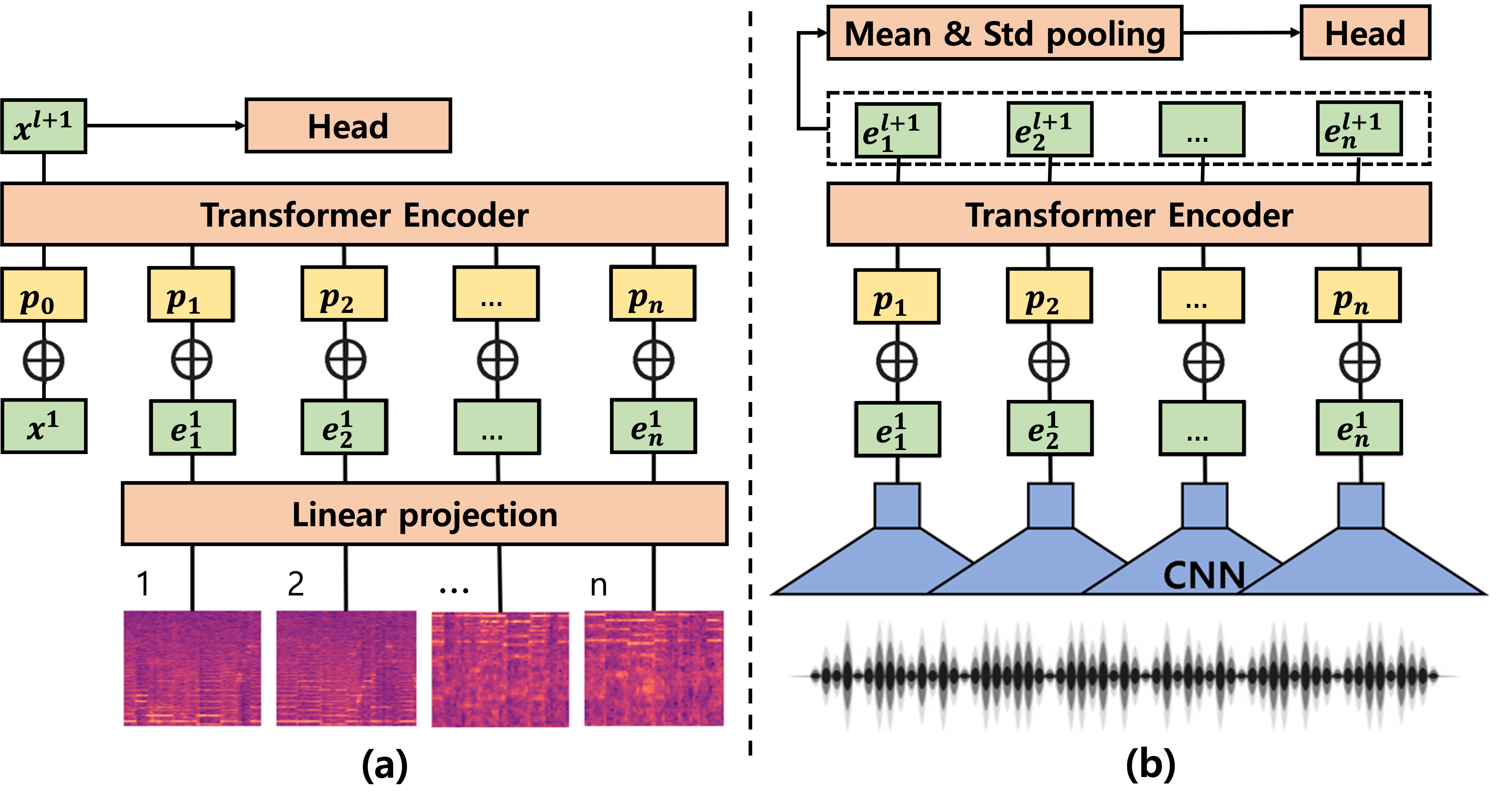}
    \vspace{-2em}
    \caption{The diagrams of (a) audio spectrogram transformer and (b) wav2vec 2.0. 
    ($x$: classification token, $e$: embedding, $p$: positional embedding.)}
\label{figure:pm_framework}
\end{center}
\vspace{-3em}
\end{figure}

\section{Methods}
This study aims to effectively and efficiently utilize PMs for audio downstream tasks without full fine-tuning. 
Prompt-based learning and adapter approaches differ in terms of how PM is used, but both have a similar purpose: steering the model results for downstream tasks while avoiding changes to the original parameters. 
Leveraging the superior transferability of these methods, we devised an \textbf{integrated parameter-efficient tuning (IPET)} framework combining prompt-based learning and adapter approaches. 
To prove the effectiveness of the proposed method, we exploited two general-purpose audio PMs with different characteristics: \textbf{audio spectrogram transformer (AST)} \cite{gong2021ast} and \textbf{wav2vec 2.0 (W2V2)}\cite{baevski2020wav2vec}. 
This section introduces the backbone PMs and the proposed IPET framework. 

\subsection{Pre-trained models}
The AST is a transformer-based classification model trained to categorize sounds such as human, animal, and things using AudioSet \cite{gemmeke2017audio}, a large dataset of about 1TB, and has demonstrated promising performance in audio classification tasks. 
We considered the AST as a supervised general-purpose audio PM from the viewpoint that it was pre-trained to classify large amounts of general audio data. 
Figure \ref{figure:pm_framework} (a) describes the structure of AST, and the pipeline is as follows: 
First, we divide the input spectrogram into $n$ patches of size $16 \times 16$. 
Then, each patch is projected as a feature embedding $E^1=\{e_1^1, ..., e_n^1 \in \mathbb{R}^{768}\}$ through a linear projection layer. 
The positional embeddings $P = \{p_0, ..., p_n \in \mathbb{R}^{768}\}$ are added to $E^1$ and a classification token ($x^1 \in \mathbb{R}^{768}$), which are fed to the transformer encoder consisting of $l$ layers. 
Finally, the processed classification token ($x^{l+1}$) output from the transformer encoder is converted into logit for classification via the task-specific liner layer ($Head$). 

The W2V2 is constructed to extract speech representation from raw waveforms by learning to predict masked speech token using the LibriSpeech corpus \cite{panayotov2015librispeech}. 
This model achieved state-of-the-art performance for automatic speech recognition, and also exhibited remarkable adaptability for other speech sub-tasks such as emotion recognition and speaker verification \cite{panayotov2015librispeech, martin2022vicomtech, vaessen2022fine}. 
Based on these observations, we exploited the W2V2 as a self-supervised general-purpose speech PM. 
The structure of W2V2 is depicted in Figure \ref{figure:pm_framework} (b). 
Note that several self-supervised training modules, such as quantization and masking, are omitted in Figure \ref{figure:pm_framework} (b) because we only utilized the pre-trained W2V2 model for downstream classification tasks. 
The W2V2 divides the raw waveform into segments and converts each to latent speech representation $E^1$ using CNN. 
Then, positional embeddings are added to $E^1$ before the transformer encoder. 
The subsequent transformer encoder outputs processed embeddings, and we apply the mean and standard deviation pooling to the final embeddings on the time axis as \cite{vaessen2022fine}. 
The pooled embedding is projected to logit through the last linear layer to perform the classification task.


\begin{figure}[t!]
\begin{center}
    \centering
    \includegraphics[width=6.5cm]{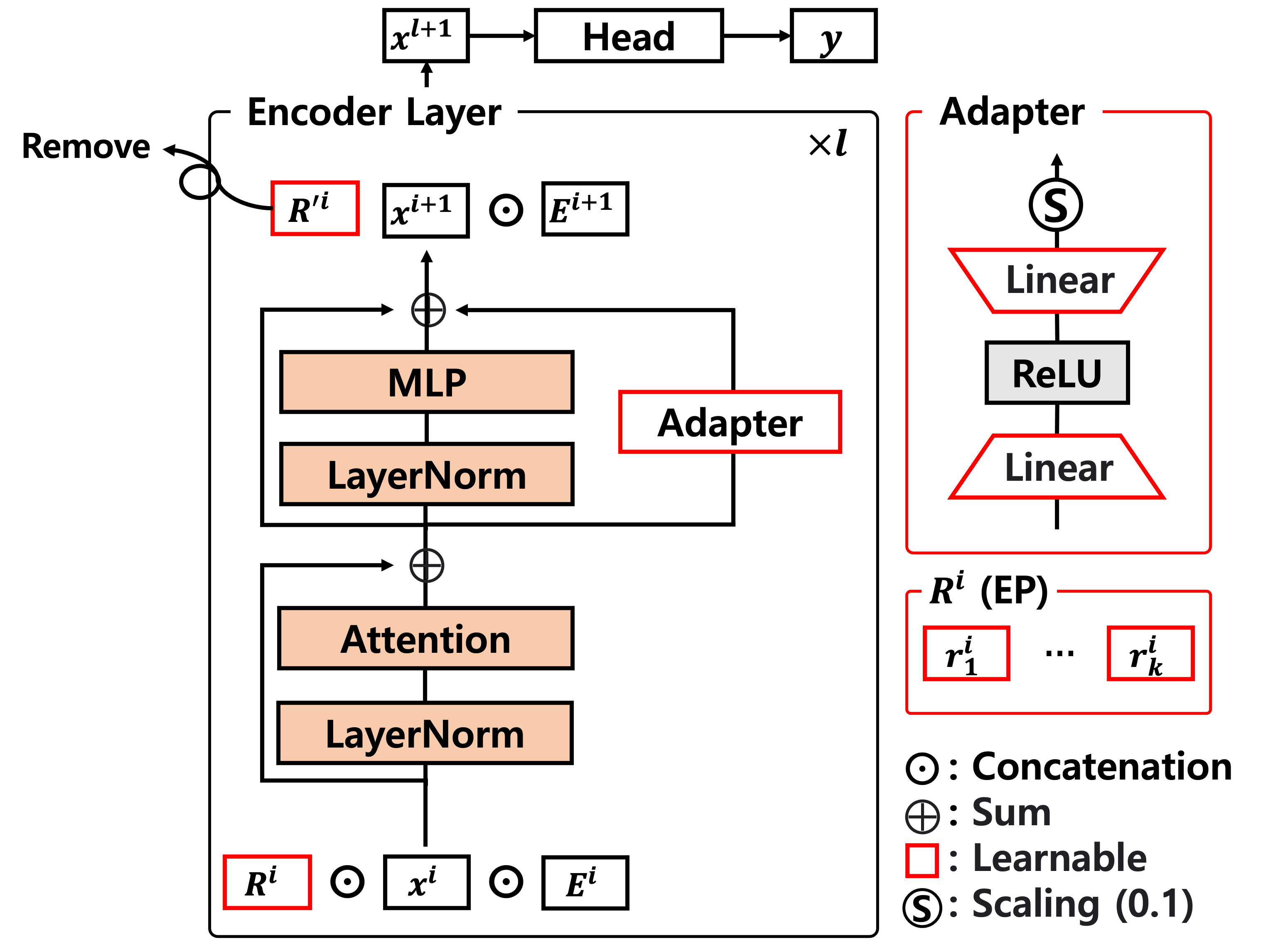}
    \vspace{-0.5em}
    \caption{The structure of proposed integrated parameter-efficient tuning. (EP: embedding prompt.)}
\label{figure:proposed}
\end{center}
\vspace{-2em}
\end{figure}

\subsection{Integrated parameter-efficient tuning (IPET)}
In this research, we explored existing transfer learning techniques and proposed the IPET framework. 
The IPET is a strategy that combines two parameter-efficient tuning methods: embedding prompt (EP) \cite{DBLP:journals/corr/abs-2203-12119} and adapter \cite{rebuffi2017learning}. 
We describe each approach based on the AST model. 
As shown in Figure \ref{figure:proposed}, \textbf{EP} method concatenates embedding prompts ($R^i = \{r_1^i, ..., r_k^i \in \mathbb{R}^{768}\}$), which are learnable parameters, to each layer's input ($x^i$ and $E^i$), and prompted embeddings are fed to the $i$-th encoder layer ($f^i$); 
where $k$ is the number of EPs. 
Then, the processed prompt ($R'^{i}$) from each preceding layer is removed and then a new prompt ($R^{i+1}$) is concatenated. 
Finally, the classification token ($x^{l+1}$) output from the last encoder layer is fed to the task-specific linear layer ($Head$) to be converted to class probability distribution ($y$). 
Only the parameters of the EP and last linear layer are trained for downstream tasks. 
\begin{equation}
\begin{aligned}
    &[\texttt{\underline{{ }{ }}} , x^{i+1}, E^{i+1}] = f^i([R^{i}, x^{i}, E^{i}]),  \quad i = 1, 2, ..., l& \\
    &y = Head(x^{l+1})&
\end{aligned}
\end{equation}
\textbf{Adapter} is a transfer learning method directly applied to a transformer encoder structure of the PM. 
In Figure \ref{figure:proposed}, it adds a thin module consisting of two linear layers, $W^i_{down}\in \mathbb{R}^{768 \times h}$ and $W^i_{up} \in \mathbb{R}^{h \times 768}$, a ReLU activation function, and a scaling factor $s$ to each encoder layer of transformer; 
where $h$ is the hidden node size of the adapter. 
The adapter receives the processed data ($B^i$) from the layer normalization ($Norm$) and attention ($Att$) layers and adds the adapted feature ($A^i$) to the output of the following multi-layer perceptron ($MLP$). 
Only the parameters of the adapter and last linear layer are trained for downstream tasks. 
\begin{equation}
\begin{aligned}
    &B^i = Att(Norm([x^i, E^i])) + [x^i, E^i],  \quad i = 1, 2, ..., l& \\
    &A^i = s \times (ReLU(B^i \cdot W^i_{down}) \cdot W^i_{up}), \quad s=0.1& \\
    &[x^{i+1}, E^{i+1}] = MLP(Norm(B^i)) + B^i + A^i& \\
    &y = Head(x^{l+1})&
\end{aligned}
\end{equation}
In summary, we proposed IPET, an effective transfer learning framework, by fusing an EP that changes the distribution of embedding data and an adapter that directly manipulates the model's behavior. 
Therefore, IPET can effectively steer the PM's results for downstream tasks. 

We compared the IPET with two different tuning methods. 
\textbf{Linear probing (LP)}, an intuitive transfer learning method, only trains the final linear layer for sub-tasks without modifying the parameters of the PM. 
\textbf{Input prompt (IP)} \cite{bahng2022exploring} applies prompts only to the input level, not to each embedding. 
Thus, it adds learnable parameters to the input, such as waveform or spectrogram.

\begin{table}[!ht]
\label{table:dataset}

\caption{
    Description of the downstream tasks, datasets, and the corresponding evaluation setup. 
    (SEC: sound event classification, MGC: music genre classification, KS: keyword spotting, SV: speaker verification.) 
    In the Setup column, the number refers the number of folds, and Split means the train, validation, and test official scenarios. 
}
\vspace{-0.5em}
\centering
\resizebox{0.95\linewidth}{!}{%
\label{table:dataset}
\begin{tabular}{c|c|c|c|c}
\Xhline{2\arrayrulewidth}
Task & Dataset & \# of classes & \# of samples & Setup\\ \hline
\multirow{2}{*}{SEC} & ESC50 &50 & 2,000 &5\\  
& FSD50K &200& 51,197 &Split\\ \hline 
MGC & GTZAN &10& 1,000 &10 \\ \hline
KS & Speech Commands V2 &35& 105,829 &Split\\ \hline
SV & VoxCeleb1 & 1,251& 153,516 &Split\\ 
\Xhline{2\arrayrulewidth}
\end{tabular}
}
\vspace{-1em}
\end{table}

\section{Experimental settings}
\label{sec:exp}
\subsection{Downstream tasks and datasets}
\label{sec:datasets}
We used a total of five datasets for four downstream tasks. 
Table \ref{table:dataset} shows the details of each dataset and sub-task. 
Sound event classification (SEC) is the task of identifying events in an input sound data; the ESC50 \cite{piczak2015esc} and FSD50K \cite{fonseca2021fsd50k} datasets were used. 
The evaluation metric for the ESC50 is accuracy (Acc), and the evaluation metric for FSD50K is mean precision (mAP), as it is a multi-label classification task. 
Music genre classification (MGC) aims to categorize musical genres, such as jazz, pop, and classical, of the input waveform. 
The GTZAN\cite{sturm2013gtzan} dataset was exploited for this task and evaluated by the Acc metric. 
Keyword spotting (KS) is the task of detecting the target keywords in given utterances, and we employed Speech Commands V2 \cite{warden2018speech} dataset. 
It was also evaluated using the Acc metric. 
Finally, speaker verification (SV) is a task of verifying whether the speaker of the input utterance is identical to the target speaker, and we utilized VoxCeleb1 \cite{Nagrani17} dataset. 
Unlike the aforementioned classification task, after training the model to classify the speakers in the dataset, we performed verification task based on the cosine similarity score between speaker embeddings extracted from utterances through the last hidden layer. 
We used the equal error rate (EER) as an evaluation metric for the SV task. 

\subsection{Implementation details}
\label{sec:details}
We compared the performance of fine-tuning, LP, IP, EP, adapter, and the proposed IPET. 
We used pre-trained AST and W2V2 provided by each official Github. 
AST ingests a 128-dimensional log Mel-spectrogram with a sampling rate of 16KHz, 25-ms hamming window, and 10-ms hop size. 
Meanwhile, W2V2 directly employs the raw waveform as input. 
All experiments were trained for 30 epochs in AST and 40 epochs in W2V2, and we used Adam optimizer. 
The remaining hyperparameters such as batch size, learning rate, and data augmentation were set differently for each experiment. 
Due to space constraints of the paper, please refer to our Github \footnote{https://github.com/wngh1187/IPET} for detailed settings and experimental code.

\section{Results and Discussion}
\label{sec:results}

\begin{table*}[th]
\caption{
    Experimental results of various transfer learning techniques for several audio downstream tasks. 
    The benchmark and fine-tuning (FT) are included for comparison. 
    We also describe the percentage of trainable parameters of each model in parentheses. 
    (LP: linear probing, IP: input prompt, EP: embedding prompt, IPET: integrated parameter-efficient tuning (proposed method).) 
}
\vspace{-0.5em}
\centering
\resizebox{0.95\linewidth}{!}{%
\label{table:main}
\begin{tabular}{c c c |c c c c c}
\Xhline{2\arrayrulewidth}
\multirow{3}{*}{\makecell[c]{Backbone\\PM}} & \multirow{3}{*}{\makecell[c]{Tuning\\Method}} & \multirow{3}{*}{\makecell[c]{Avg.\\ Trainable \\ Params (M)}} & \multicolumn{2}{c}{Sound event classification} & Music genre classification & Keyword spotting & Speaker verification \\
\cline{4-8}
 & & & ESC50 & FSD50K & GTZAN & Speech Commands V2 & VoxCeleb1\\
& & &(Acc $\uparrow$) & (mAP $\uparrow$) & (Acc $\uparrow$) & (Acc $\uparrow$) & (EER $\downarrow$)\\ \hline
\multicolumn{2}{c}{Benchmark}& - & 95.7 \cite{gong2021ast} & 0.641 \cite{turian2022hear} & 90.9 \cite{DBLP:journals/prl/NanniCLSB17} & 98.11 \cite{gong2021ast} & 4.35 \cite{wang2022multi}\\
\Xhline{1\arrayrulewidth}
\multirow{6}{*}{AST} & FT & 87.47 (100\%) & 95.35 & 0.631 & 90.2 & \textbf{98.12} & \textbf{11.74}\\
& LP & 0.23 (0.27\%) & 94.95 & 0.611 & 86.3 & 68.68 & 28.05\\
& IP & 0.24 (0.27\%) & 95.05 & 0.616 & 87.1 & 79.37 & 24.06\\
& EP & 0.55 (0.63\%) & 96.05 & 0.63 & 89 & 93.72 & 18.33\\
& Adapter & 1.04 (1.17\%) & 96.35 & 0.639 & 89.9 & 96.3 &12.8\\
& IPET & 0.99 (1.12\%) & \textbf{96.45} & \textbf{0.641} & \textbf{90.8} & 96.12 &12.17\\
\Xhline{1\arrayrulewidth}
\multirow{6}{*}{W2V2} & FT & 94.52 (100\%) & \textbf{76.75} & \textbf{0.535} & \textbf{81.8} & 97.88 & 4.02 \\
& LP & 0.19 (0.2\%) & 49.5 & 0.228 & 67.8 & 82.50 & 14.51 \\
& IP & 0.20 (0.21\%) & 52.05 & 0.252 & 72.7 & 88.25 & 14.29 \\
& EP & 0.71 (0.74\%) & 63.85 & 0.37 & 74.9 & 97.56 & 5.12 \\
& Adapter & 1.75 (1.81\%) & 67 & 0.429 & 76.3 & 97.74 & 3.9 \\
& IPET & 2.85 (2.93\%) & 71.6 & 0.447 & 78.7 & \textbf{98.27} & \textbf{3.34} \\
\Xhline{2\arrayrulewidth}
\end{tabular}
}
\vspace{-1em}
\end{table*}

Table \ref{table:main} displays the results of applying various transfer learning techniques and the proposed IPET to both AST and W2V2 PMs for several audio downstream tasks. 
Models that achieved reliable performance for each task were adopted as benchmarks, and the number of trainable parameters for each method were calculated based on the best version. 
We analyzed the results in three aspects: \textbf{(1) previous methods vs. proposed IPET}, \textbf{(2) matched vs. mismatched domains}, \textbf{(3) supervised vs. self-supervised PMs}. 
We also discussed the limitations of this paper and future works. 

\subsection{Transfer learning and IPET methods}
We evaluated each technique based on its overall performance for sub-tasks. 
Although fine-tuning (FT) was a powerful transfer learning solution compared to the benchmark performance, it required re-training a large number of parameters. 
On the other hand, LP only needed to adjust the last classification layer, but did not consistently derive satisfactory results. 
It was demonstrated that prompt-based learning methods are better than LP and that prompting in the embedding space is more effective than in the input space. 
Also, the adapter method exhibited enhanced adaptation ability by applying it directly inside the PM compared to prompt-based learning. 
The proposed IPET achieved superior performance compared to the existing tuning techniques, comparable to or surpassing benchmark performance under matched domain conditions (refer to Section \ref{sub:inout}). 
Based on these results, we concluded that the IPET can appropriately shift the embedding distribution through the embedding prompt, while also adjusting the model results directly through the adapter. 

\begin{figure}[t]
\begin{center}
    \centering
    \includegraphics[width=\linewidth]{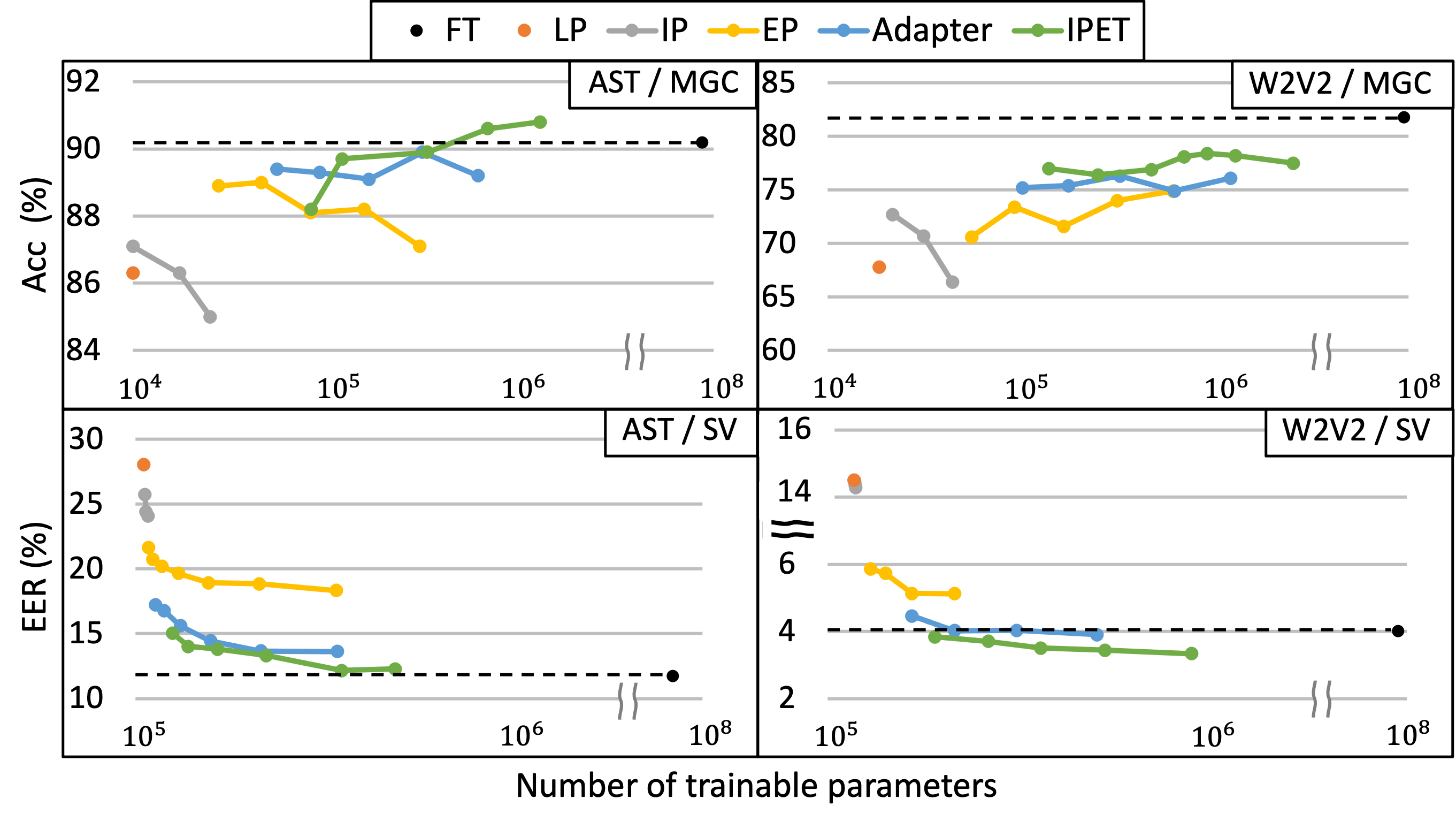}
    \vspace{-2em}
    \caption{Comparison of performance according to parameter scaling for various transfer learning methods and proposed IPET. 
    For intuitive expression, FT results are plotted with dotted lines. 
    (Left: using AST; Right: using W2V2; Top: on music genre classification (MGC); Bottom: on speaker verification (SV).)}
\vspace{-2em}
\label{figure:chart}
\end{center}
\vspace{-0.5em}
\end{figure}

Furthermore, to prove the superiority of the IPET, we observed the performance of transfer learning methods by varying the number of parameters for both PMs on the MGC and SV tasks. 
As shown in Figure \ref{figure:chart}, the IPET framework achieved superior results based on a similar number of weights compared to other methods. 
In addition, the IPET performance improved stably (more than conventional approaches) as the scale of parameter set increased.

\subsection{Matched and mismatched domains results}
\label{sub:inout}
We assumed a matched domain condition if the labels or data types between the downstream task's dataset and the PM's pre-training dataset are similar, and mismatched domain otherwise. 
The AudioSet, the training dataset of AST, contains specific labels for a wide range of sound event, including sounds of things and music. 
Therefore, even simple transfer learning such as LP showed tolerable results to FT in a matched domain environment with similar or overlapping labels, such as ESC50, FSD50K, and GTZAN datasets in AST experiments. 
However, since speech data included in AudioSet is abstractly labeled (e.g., male speech, female speech, and child speech), the LP scheme suffered from severe performance degradation for mismatched domain examples, keyword and speaker identity data (Speech Commands V2 and Voxceleb1). 
On the other hand, W2V2 is a self-supervised model that learned extracting speech representations using LibriSpeech dataset. 
Thus, it delivered inferior performance, even with FT as well as LP, in mismatched domain tasks (SEC, MGC). 
Moreover, because W2V2 did not directly learn the characteristics of downstream tasks, the LP method did not achieve sufficient performance in matched domain data tasks such as KS or SV. 
Under these scenarios and conditions, the proposed IPET showed improved transferability in the matched domain condition and more stable results in the mismatched domain than other transfer learning techniques. 
Through these results, we interpreted that the proposed IPET technique is effective and reliable in both matched and mismatched domains.

\subsection{Supervised and self-supervised PMs}
As mentioned above, the proposed IPET demonstrated superiority regardless of the PM's pre-training learning strategy. 
The AST, supervised PM, exhibited relatively satisfactory performance with only LP in the matched domain tasks, and the proposed method was able to further improve the adaptability of the model. 
Based on this tendency, we hypothesized that IPET could be considered as a kind of domain adaptation that encourages the utilization of previously supervised knowledge in other domains. 
Therefore, we expect that the proposed technique can also be applied to other domain adaptation scenarios. 
Meanwhile, the self-supervised models have generally been employed as a feature extractor to be applied for the target task \cite{baevski2020wav2vec}. 
In other words, the self-supervised PMs were exploited in conjunction with a task-specific back-end model. 
On the contrary, this study directly applied W2V2 to downstream tasks in an end-to-end manner, and the IPET technique achieved remarkable performances without a deep back-end module in the matched domain condition. 
Through these results, we confirmed that IPET is a method that can effectively use the ability of the self-supervised model. 

\subsection{Limitations and future works} 
We compared several parameter-efficient tuning methods for the matched and mismatched domains using two PMs with different characteristics. 
Through experiments and analysis, this paper demonstrated the effectiveness of the proposed IPET. 
Simultaneously, we identified limitations to the present study. 
First, we considered only the transformer-based PM and performed the classification-focused downstream tasks. 
In addition, though IPET delivered relatively stable performance compared to other tuning methods in the mismatched domain condition, the performance itself is not satisfactory. 
Therefore, as a future work, we will further study a variant of IPET applicable to existing CNN-type models, and prove the efficacy of the proposed framework using an official audio benchmark dataset including more diverse sub-tasks \cite{turian2022hear}. 
Moreover, we will verify the transferability of the IPET using other supervised PMs that can encompass various sub-tasks or self-supervised PMs that have learned from comprehensive audio datasets.

\section{Conclusion}
\label{sec:con}
We proposed the IPET framework to effectively and efficiently utilize pre-trained models for audio downstream tasks. 
We compared the IPET with various transfer learning approaches by applying them to two PMs with different characteristics, AST and W2V2. 
Experiment results showed IPET to be more effective than the existing parameter-efficient tuning technique. 
In addition, under the matched domain condition, where the domain of the PM's pre-training dataset and the sub-task's data domain were similar, the proposed framework outperformed the benchmark or showed comparable results. 
Furthermore, IPET exhibited satisfactory transferability regardless of the PM's training manner.
However, we also identified some drawbacks of the proposed method and our study. 
As a future work, we will overcome the aforementioned limitations. 
Finally, we hope that our work will ultimately help with the study of transfer learning in the audio domain community. 

\bibliographystyle{IEEEbib}
\bibliography{mybib}

\end{document}